\documentclass[twocolumn]{aastex631}

\def\gtsima{$\; \buildrel > \over\sim \;$}
\def\gtsim{\lower.5ex\hbox{\gtsima}}
\def\ltsima{$\; \buildrel < \over\sim \;$}
\def\ltsim{\lower.5ex\hbox{\ltsima}}

\shorttitle{DSNB with a New GCE Model}
\shortauthors{Ashida, Nakazato, and Tsujimoto}
\graphicspath{{./}{figure/}}

\begin{document}

\title{Diffuse Neutrino Flux Based on the Rates of Core-collapse Supernovae and Black Hole Formation Deduced from a Novel Galactic Chemical Evolution Model}

\author[0000-0003-4136-2086]{Yosuke Ashida}
\affiliation{Department of Physics and Astronomy, University of Utah, Salt Lake City, UT 84112, USA}
\affiliation{Department of Physics and Wisconsin IceCube Particle Astrophysics Center, University of Wisconsin-Madison, Madison, WI 53706, USA}

\author[0000-0001-6330-1685]{Ken'ichiro Nakazato}
\affiliation{Faculty of Arts and Science, Kyushu University, Fukuoka 819-0395, Japan}

\author[0000-0002-9397-3658]{Takuji Tsujimoto}
\affiliation{National Astronomical Observatory of Japan, Mitaka, Tokyo 181-8588, Japan}




\begin{abstract}

Fluxes of the diffuse supernova neutrino background (DSNB) are calculated based on a new modeling of galactic chemical evolution, where a variable stellar initial mass function (IMF) depending on the galaxy type is introduced and black hole (BH) formation from the failed supernova is considered for progenitors heavier than 18$M_{\odot}$. 
The flux calculations are performed for different combinations of the star formation rate, nuclear equation of state, and neutrino mass hierarchy to examine the systematic effects from these factors. 
In any case, our new model predicts the enhanced DSNB $\bar{\nu}_{e}$ flux at $E_\nu \gtsim 30$~MeV and $E_\nu \ltsim 10$~MeV due to more frequent BH formation and a larger core-collapse rate at high redshifts in early-type galaxies, respectively. 
Event rate spectra of the DSNB $\bar{\nu}_{e}$ at a detector from the new model are shown and the detectability at water-based Cherenkov detectors, Super-Kamiokande with a gadolinium dissolution and Hyper-Kamiokande, is discussed. 
In order to investigate the impacts of the assumptions in the new model, we prepare alternative models, based on different IMF forms and treatments of BH formation, and estimate the discrimination capabilities between the new and alternative models at these detectors. 

\end{abstract}



\section{Introduction} \label{sec:intro}

\begin{figure}[htbp]
  \begin{center}
    \includegraphics[clip,width=8.6cm]{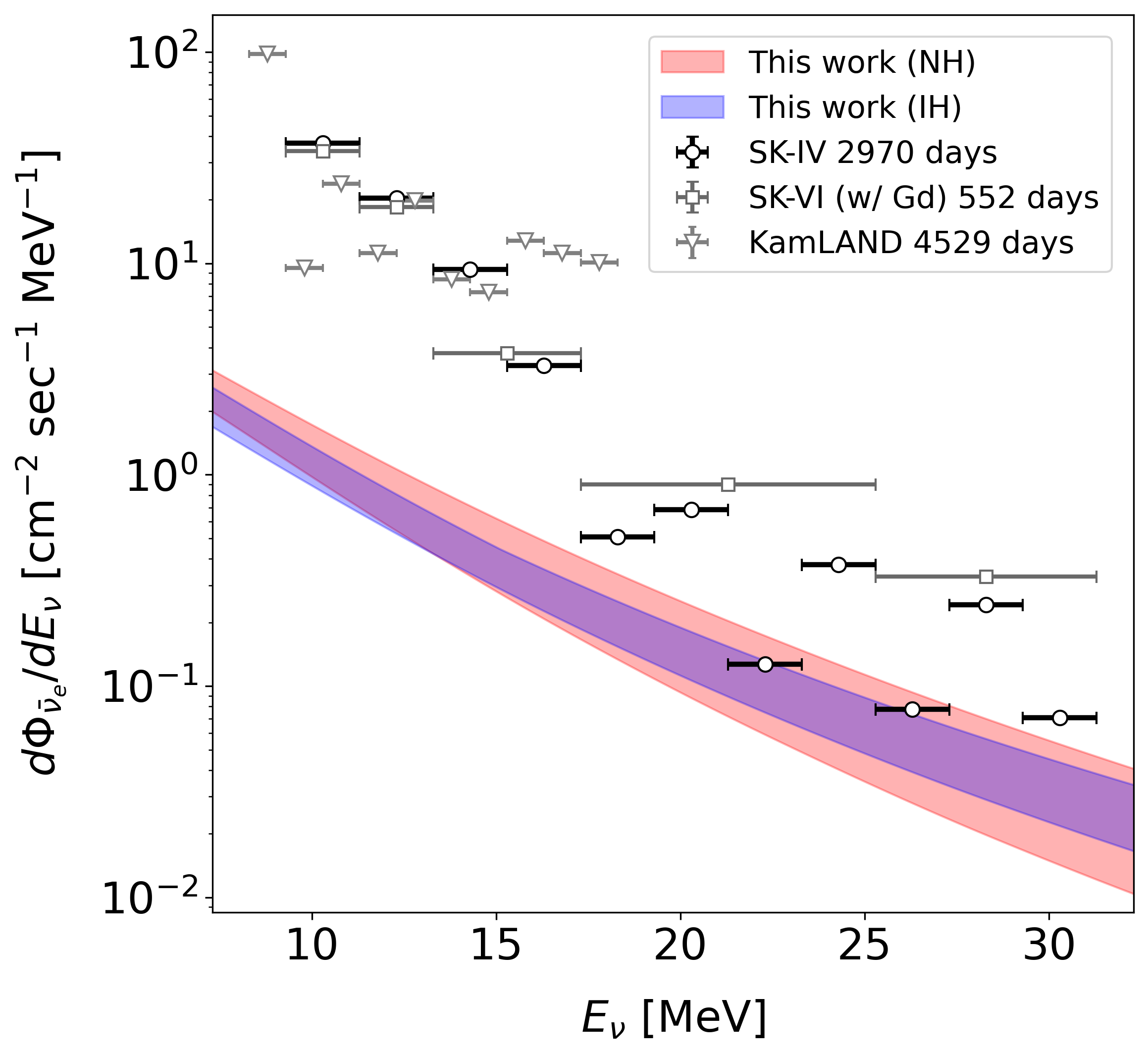}
  \end{center}
  \vspace{-10truept}
  \caption{Prediction bands of the DSNB $\bar\nu_e$ flux from this study compared with the 90\% confidence level upper bounds on the extraterrestrial $\bar\nu_e$ flux from the latest analyses at SK with pure water~\citep[][]{2021PhRvD.104l2002A} and gadolinium~\citep[][]{2023ApJ...951L..27H} and KamLAND~\citep[][]{2022ApJ...925...14A}. The bands cover the predictions with different choices of the SFR~\citep[][]{2006ApJ...651..142H,2014ARA&A..52..415M} and the nuclear EOS~\citep[][]{1991NuPhA.535..331L,2011ApJS..197...20S,2017NuPhA.961...78T}. The red (blue) band corresponds to the case with normal (inverted) neutrino mass hierarchy.}
  \label{fig:fluxcomp}
  \end{figure}

A significantly massive star causes the gravitational collapse of its core at the end of its evolution, releasing a large fraction of the binding energy as neutrinos. 
These neutrinos would provide valuable information about the deep cores of stars that are not optically accessible. 
In particular, neutrinos from the core collapse of massive stars over cosmic history have accumulated to form the diffuse supernova neutrino background (DSNB), or they are referred to as supernova relic neutrinos, and they are expected to provide knowledge of not only collapsing stars, but also cosmic chemical evolution, etc. 
Impacts of physical and astrophysical factors on DSNB have been studied actively in many theoretical works~\citep[e.g.,][]{1995APh.....3..367T,1996ApJ...460..303T,1997APh.....7..137H,1997APh.....7..125M,2000PhRvD..62d3001K,2003APh....18..307A,2009PhRvL.102w1101L,2009PhRvD..79h3013H,2010PhRvD..81e3002G,2013PhRvD..88h3012N,2015ApJ...804...75N,2018MNRAS.475.1363H,2020MPLA...3530011M,2021ApJ...909..169K,2021JCAP...05..011T,2021PhRvD.103d3003H,2022PhRvD.106d3026E,2022ApJ...937...30A,2022MNRAS.517.2471Z,2023ApJ...946...69A,2023ApJ...950...29A}.
In the experimental field, an observation of the DSNB flux has yet to be accomplished; however, a recent search at the Super-Kamiokande (SK) water Cherenkov detector has excluded some optimistic models and placed upper bounds on the DSNB $\bar{\nu}_{e}$ flux close to many theoretical predictions, within a factor of 10~\citep[][]{2021PhRvD.104l2002A}. 
Figure~\ref{fig:fluxcomp} shows the latest experimental upper bounds as well as the new model predictions in this paper. 
Recently, SK was upgraded with a gadolinium dissolution, as SK-Gd, for the purpose of the first detection of DSNB~\citep[][]{2004PhRvL..93q1101B,2022NIMPA102766248A} and the first $\bar{\nu}_{e}$ search result was shown in \citet{2023ApJ...951L..27H}. 
Furthermore, another water Cherenkov detector with a megaton scale, Hyper-Kamiokande (HK), will be constructed in the next decade~\citep[][]{2018arXiv180504163H,2018PTEP.2018f3C01H}. 
It is expected that HK will be able not only to confirm the existence of the DSNB, but also to determine the flux shape, which should be a reflection of the underlying physics. 
In addition to the water-based Cherenkov detector, liquid scintillator experiments have been running for the $\bar{\nu}_{e}$ search, and the KamLAND experiment achieves the best sensitivity at low energies~\citep[][]{2022ApJ...925...14A}. 
In the next few decades, many types of detectors ---including liquid scintillators and argon-/xenon-based detectors--- are planned for construction and expected to be sensitive enough to the DSNB flux~\citep[see][]{2017JCAP...11..031P,2018JCAP...05..066M,2020PhRvD.102l3012D,2021PhRvD.103b3021S,2022Univ....8..181L,2022PhRvD.105d3008S}. 
Measuring at multiple types of detectors could help to determine the flux over the wide energy range and therefore provide more reliable and valuable insights into the physics behind the DSNB. 

For the prediction of the DSNB flux, the frequency of the neutrino emitters, i.e., core-collapse supernovae (CCSNe) and black hole (BH) formation from the failed SN, as a function of time (or redshift) is crucial. 
A conventional approach to deducing the rate of CCSNe ($R_{\rm CCSN}$) is to scale the observed cosmic star formation rate (SFR), based on the assumption that the death rate of massive stars is proportional to the SFR, so as to match the measured relation between the CCSN rate and redshift. 
In this approach, the stellar initial mass function (IMF) is assumed to be universal in the Universe \citep[see][]{2010ARA&A..48..339B}. 
Recently, \citet{2023MNRAS.518.3475T} has argued against the universality of the IMF from the aspect of galactic chemical evolution, suggesting that the IMF could vary depending on the types of galaxies \citep[see][]{2018PASA...35...39H}; the early-type galaxy ---represented by the elliptical galaxy--- that is formed in a bursty star formation should have an IMF generating more massive stars and thereby ejecting more heavy elements than that of a late-type galaxy like our own. 
\citet{2023MNRAS.518.3475T} also shows that the predicted redshift evolution of $R_{\rm CCSN}$ with a variable IMF among galaxies gives a much better fit to the measured rates\footnote{The models with a variable IMF are shown to succeed in explaining the observed large contrast in the CCSN rate between the present and $z\approx1$. However, \citet{2022MNRAS.517.2471Z} predict a conversely smaller contrast using a model with a variable IMF compared to one with a nonvariable IMF (see Section~\ref{sec:discuss}).}. 
The central argument about the variable IMF comes from a recent finding that the upper bound on the mass of CCSN progenitor stars is as small as 18$M_\odot$ \citep[][]{2009ARA&A..47...63S,2015PASA...32...16S,2016ApJ...821...38S,2021ApJ...909..169K}. 
Accordingly, the proposed IMF predicts a high frequency of BH formation as the fate of core-collapse stars with a mass larger than $18 M_\odot$, and thus calls for a careful consideration of the rate of BH formation ($R_{\rm BH})$, which has been ignored in the majority of existing models, including the ones with a variable IMF \citep[e.g.,][]{2023ApJ...946...69A}, when calculating the DSNB flux. 
Therefore, the newly estimated $R_{\rm CCSN} (z)$ and $R_{\rm BH} (z)$, as demanded by galactic chemical evolution, would significantly lift the accuracy of the DSNB flux prediction.

This paper shows the DSNB flux based on the new galactic chemical evolution model and also discusses the experimental detectability. 
The rest of the article is structured as follows. 
First, we describe modeling of the DSNB flux in Section~\ref{sec:model}.
We then show event rate spectra at the water Cherenkov detector from our new model in Section~\ref{sec:exp}. 
In Section~\ref{sec:sens}, we estimate experimental sensitivities at SK-Gd and HK. 
Discussion is provided in Section~\ref{sec:discuss} before conclusion is given in Section~\ref{sec:concl}.

\section{DSNB Model} \label{sec:model}

\subsection{CCSN and Failed SN Rates with a Variable IMF}

The CCSN rate, $R_{\rm CCSN}(z)$, is calculated by converting from the observationally estimated cosmic SFR ($\psi$) including representative ones \citep{2006ApJ...651..142H, 2014ARA&A..52..415M} via a scale factor, $k_{\rm CCSN}$, of massive stars that explode as CCSNe per unit mass of IMF, which is related by $R_{\rm CCSN}(z)=k_{\rm CCSN}(z)\cdot\psi(z)$. 
Here, the units of $R_{\rm CCSN}$ and $\psi$ are yr$^{-1}$ Mpc$^{-3}$ and $M_{\odot}$ yr$^{-1}$ Mpc$^{-3}$, respectively. 
In our calculations, $k_{\rm CCSN}$ varies in accordance with $z$ since the $\psi(z)$ value is a composite of the SFR contributed from different types of galaxies in which the star formation history differs and the IMF is assumed to change. 
Thus, the type $j$ galaxy has a unique value of $k_{\rm CCSN}$, i.e., $k_{{\rm CCSN},j}$.

In order to evaluate $k_{\rm CCSN} (z)$, we classify galaxies into five groups: spheroids (E/S0) and four classes of spiral galaxies (Sab, Sbc, Scd, and Sdm). 
Given their star formation histories, the relative contribution to $\psi (z)$ from each galaxy type is calculated by weighting with its relative proportion and mass-to-luminosity ratio \citep{1996ApJ...460..303T}. 
Then, $k_{\rm CCSN}(z)$ is defined as
  \begin{equation}
    k_{\rm CCSN}(z)=\sum_{j=1}^5 k_{{\rm CCSN},j}\frac{w_j \psi_j(z)}{\sum_{j=1}^5 w_j \psi_j(z)},
  \end{equation}
where $w_j$ and $\psi_j(z)$ are the individual weight and SFR for the type $j$ galaxy, respectively. 
Based on knowledge from the study of galactic chemical evolution \citep{2023MNRAS.518.3475T}, we adopt the Salpeter IMF (a slope index of a power law $x=-1.35$) for the late-type galaxies (Sbc, Scd, and Sdm) and the flat IMF ($x=-0.9$) for the early-type ones (E/S0 and Sab) to count the number of CCSNe whose progenitor masses are in the range of 8--18$M_{\odot}$. 
Given the entire mass range of 0.01--100$M_{\odot}$ with a single slope of the Salpeter IMF, $k_{\rm CCSN}$ is deduced to be $\sim$0.005$M_\odot^{-1}$. 
Considering the uncertainty in the complex mass distribution for low-mass stars ($<$1$M_{\odot}$), we adopt a slightly modified $k_{\rm CCSN}$ of 0.006$M_\odot^{-1}$ for the late-type galaxies.
In the same way, we have $k_{\rm CCSN} = 0.018$$M_\odot^{-1}$ for the flat IMF adopted for the early-type galaxies. 
Then, we finally obtain $R_{\rm CCSN} (z)$ by combining $k_{\rm CCSN} (z)$ with the measured $\psi (z)$.
In this study, we adopt the $\psi (z)$ of \citet{2006ApJ...651..142H} and \citet{2014ARA&A..52..415M}, which are referred to as HB06 and MD14, respectively.

The BH formation rate, $R_{\rm BH}(z)$, is deduced in the same manner, just by shifting the mass range of progenitor stars from 8--18$M_{\odot}$ for CCSNe to 18--100$M_{\odot}$ for failed SNe forming BHs. 

In the present study, we show the DSNB flux based on a variable IMF and the BH formation for progenitors with 18--100$M_{\odot}$ as a reference model. 
In addition to it, we prepare three alternative models, depending on choices of the IMF form (variable or Salpeter) and treatment of the BH formation (a BH formed for 18--100$M_{\odot}$ progenitors or a CCSN assumed for all stellar core collapses), for comparison.
We use a special nomenclature for these models, as summarized in Table~\ref{tab:modelnomenclature} and will refer to it in the rest of this paper. 

  \begin{table}[htbp]
  \begin{center}
  \caption{Nomenclature for the DSNB model depending on choices of IMF and BH treatment. GDIMF-wBH is a reference model in this study.}
  \label{tab:modelnomenclature}
  \vspace{-8truept}
    \begin{tabular}{lll} \hline\hline 
      Name & IMF form & BH treatment \\ \hline 
      GDIMF-wBH (ref.)   & Variable & BHs for 18--100$M_{\odot}$ \\ 
      GDIMF-noBH  & Variable & No BH \\
      SalIMF-wBH  & Salpeter & BHs for 18--100$M_{\odot}$ \\
      SalIMF-noBH & Salpeter & No BH \\ \hline\hline 
    \end{tabular}
  \end{center}
  \end{table}

\subsection{Neutrino Spectra from a CCSN and a Failed SN}

The spectrum and total energy of the emitted neutrinos from the stellar core collapse depend on the properties of the compact remnants. 
As for the case that a neutron star (NS) is formed, the total energy of the emitted neutrinos is determined by the NS mass.
According to \citet{2020ApJ...896...56W}, the NS mass distribution has a central peak near $1.35M_\odot$ and the fraction of high-mass NSs at birth is far lower\footnote{While the fraction of high-mass NSs at birth is still under debate \citep[e.g.,][]{2016arXiv160501665A}, their progenitors are generally considered to be highly massive. 
For instance, \citet{2022MNRAS.517.2471Z} adopt a $27M_\odot$ progenitor for CCSNe with a high-mass NS. In contrast, in our model, progenitors heavier than $18M_\odot$ form a BH and we do not consider the existence of high-mass NSs at birth.}. 
Thus, in the present study, we adopt the neutrino spectrum of the canonical-mass NS model with $\sim$1.35$M_\odot$ in \citet{2022ApJ...937...30A}, which is denoted as $dN_{\rm CNS}(E^\prime_\nu)/dE^\prime_\nu$ in that paper, as a reference model of the neutrino spectrum from a CCSN: $dN_{\rm CCSN}(E^\prime_\nu)/dE^\prime_\nu$.
%
%
Here, $E^\prime_\nu$ is the neutrino energy at a source.
We further utilize a reference model of the neutrino spectrum in that paper for a failed SN with a BH: $dN_{\rm BH}(E^\prime_\nu)/dE^\prime_\nu$. 
In the BH formation case, the neutrino spectrum is harder than that of the case with an NS. 
More details of the reference model for the neutrino spectrum in this study are described in \citet{2022ApJ...937...30A}. We then describe the DSNB flux as
  \begin{eqnarray}
    & & \frac{d\Phi(E_\nu)}{dE_\nu}=  c\int^{z_{\rm max}}_{0}  \frac{dz}{H_0\sqrt{\Omega_{\rm m}(1+z)^3+\Omega_\Lambda}} \times \nonumber \\ & & \left[ R_{\rm CCSN}(z) \frac{dN_{\rm CCSN}(E^\prime_\nu)}{dE^\prime_\nu} + R_{\rm BH}(z) \frac{dN_{\rm BH}(E^\prime_\nu)}{dE^\prime_\nu} \right],
  \label{eq:flux}
  \end{eqnarray}
where $c$ is the speed of light and the cosmological constants are $\Omega_{\rm m}= 0.3089$, $\Omega_\Lambda= 0.6911$, and $H_0= 67.74$~km~sec$^{-1}$~Mpc$^{-1}$. 
For the source located at a redshift $z$, $E^\prime_\nu$ is related to the neutrino energy at a detector $E_\nu$ as $E^\prime_\nu=E_\nu(1+z)$. In this study, we set $z_{\rm max}=5$.

The nuclear equation of state (EOS) is also a factor that affects the neutrino emission from the stellar core collapse. 
In the present study, we investigate three models: the Togashi EOS~\citep[][]{2017NuPhA.961...78T}, the LS220 EOS~\citep[][]{1991NuPhA.535..331L}, and the Shen EOS~\citep[][]{2011ApJS..197...20S}. 
Here, we briefly review the impacts of the nuclear EOS, while the properties of our models are summarized in Table~1 of \citet{2022ApJ...937...30A} and the detailed description is given in \citet{2021PASJ...73..639N,2022ApJ...925...98N} and references therein. 
In the case of the SN leaving an NS, the total energy of the emitted neutrinos is larger for the EOS models with a smaller NS radius, because it is related to the binding energy of the remnant NS. 
Among the three models, the Togashi EOS has the smallest NS radius, followed by LS220 and Shen, in order of increasing radius. 
However, the average energy of the neutrinos is insensitive to the nuclear EOS, as well as the NS mass. 
On the other hand, in the case of the BH formation, the mass accretion continues until the mass of the central remnant grows to the point where it overwhelms the nuclear force.
Thus, the nuclear EOS affects the neutrino emission; the total energy of the emitted neutrinos is larger for the EOS models with a higher maximum mass of NSs because the time interval between the collapse and the BH formation, which corresponds to the duration of the neutrino emission, is longer. 
Note that the gravitational potential energy released from the collapse is not fully converted to neutrinos because the neutrino emission is terminated by the BH formation. 
The average energy of the emitted neutrinos is higher for the EOS models with a higher maximum mass of NSs because the heating of the collapsed core due to matter accretion continues until the BH formation. 
Among the three models, the Shen EOS has the highest maximum mass of NSs, followed by Togashi and LS220, in order of decreasing mass. 

Finally, we take into account the neutrino flavor oscillation for evaluating the $\bar\nu_e$ flux on Earth.
For this purpose, we consider the Mikheyev-Smirnov-Wolfenstein (MSW) effect\footnote{It is an open question if there is a potential effect beyond MSW in such a dense environment (see \citet{2023arXiv230111814V}).} following \citet{2022ApJ...937...30A} while a detailed description is given in \citet{2015ApJ...804...75N}. 
The resultant $\bar\nu_e$ flux depends on the neutrino mass hierarchy, and we investigate both cases of the normal hierarchy (NH) and inverted hierarchy (IH) in the present study.

\begin{figure}[htbp]
  \begin{center}
    \includegraphics[clip,width=8.6cm]{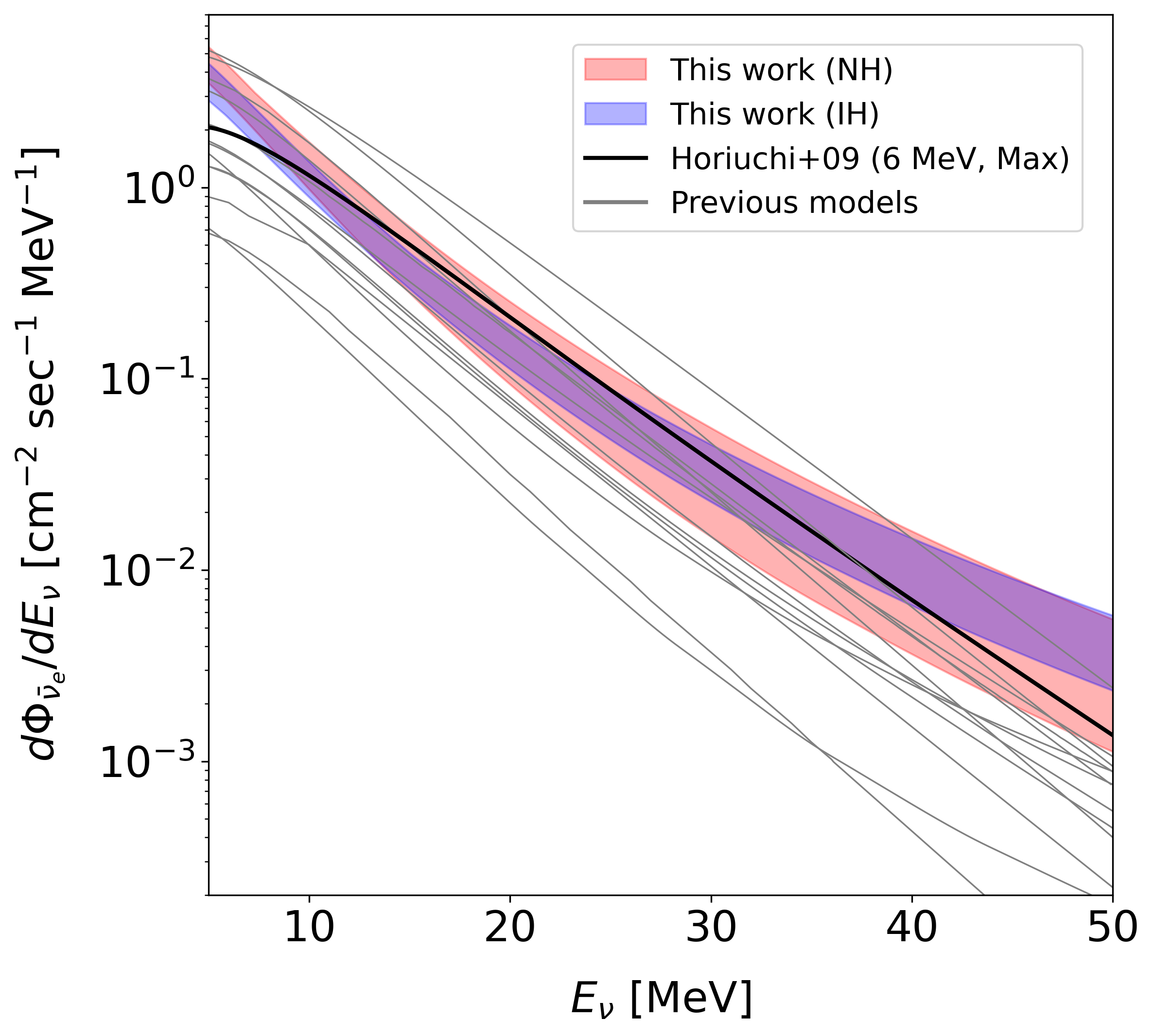}
  \end{center}
  \vspace{-10truept}
  \caption{Prediction bands of the DSNB $\bar\nu_e$ flux from this study compared with predictions from the previous theoretical works shown in Figure~1 of \citet{2021PhRvD.104l2002A}. We highlight the prediction from \citet{2009PhRvD..79h3013H} which is shown as a reference model in the SK paper~\citep[][]{2021PhRvD.104l2002A}. The color notation is the same as in Figure~\ref{fig:fluxcomp}. Refer to each publication for the model details~\citep{1995APh.....3..367T,1997APh.....7..137H,1997APh.....7..125M,2000PhRvD..62d3001K,2003APh....18..307A,2009PhRvL.102w1101L,2009PhRvD..79h3013H,2010PhRvD..81e3002G,2015ApJ...804...75N,2018MNRAS.475.1363H,2021ApJ...909..169K,2021JCAP...05..011T,2021PhRvD.103d3003H}.}
  \label{fig:fluxmodelcomp}
  \end{figure}

\begin{figure}[htbp]
  \begin{center}
    \includegraphics[clip,width=8.6cm]{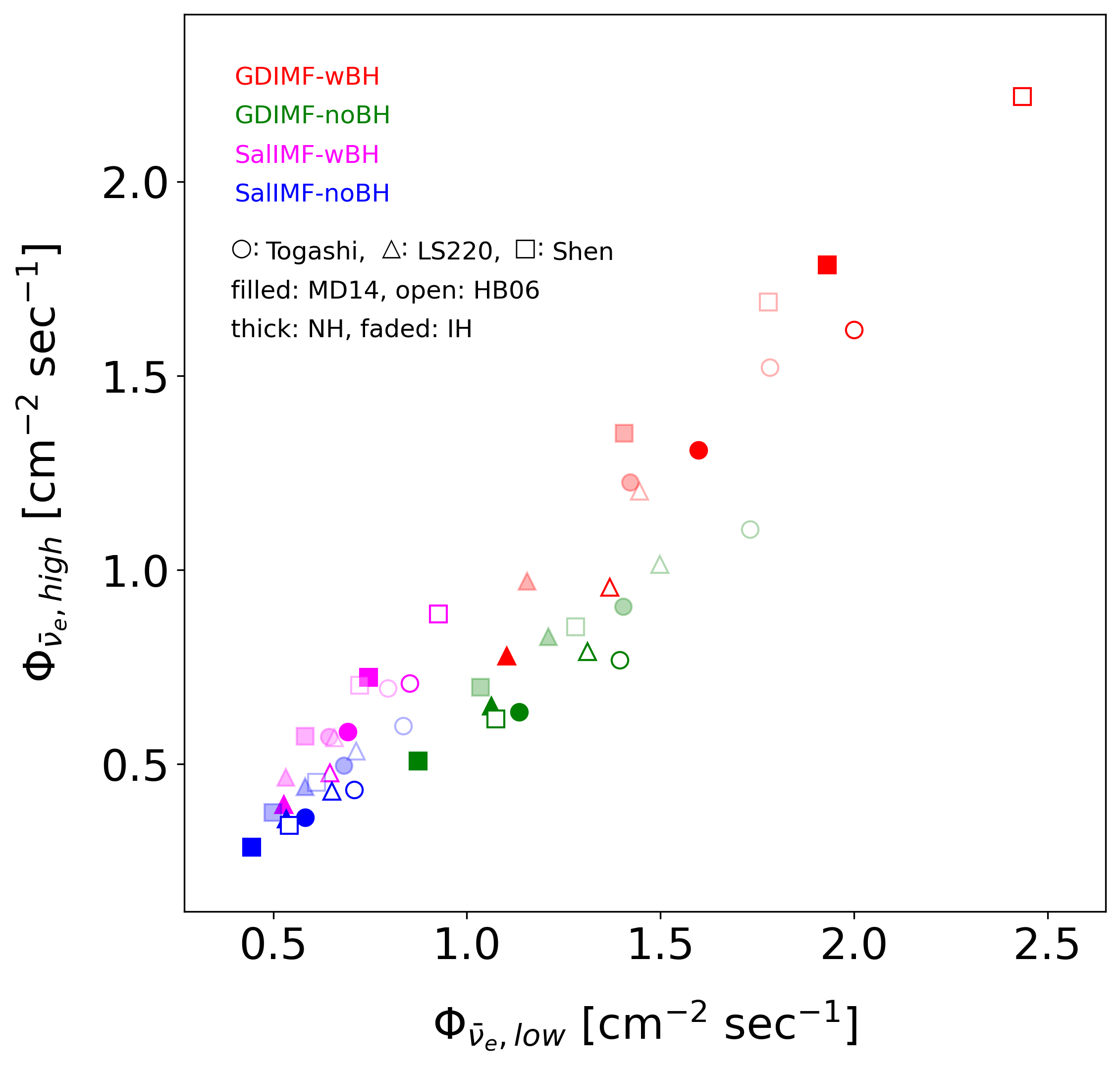}
  \end{center}
  \vspace{-10truept}
  \caption{Scatter plot on the integrated DSNB $\bar\nu_e$ fluxes for low- ($13.3\leq E_{\nu}<17.3$~MeV) and high-energy ranges ($17.3\leq E_{\nu}<31.3$~MeV), with different choices of the IMF, BH treatment, nuclear EOS, SFR, and neutrino mass hierarchy. The different colors represent the models in Table~\ref{tab:modelnomenclature}. The circle, triangle, and square symbols correspond to the Togashi, LS220, and Shen EOSs, respectively. Color-filled symbols are with MD14 and open ones with HB06 as the SFR. Normal and inverted mass hierarchy cases are drawn with thick and faded colors, respectively.}
  \label{fig:fluxintegcomp}
  \end{figure}

Figure~\ref{fig:fluxmodelcomp} shows the DSNB $\bar\nu_e$ flux predictions as bands, which cover the different SFR and EOS models, in comparison with previous theoretical studies. 
While the predictions from the previous works show substantial variation, our results fall within the range of this variation below $\sim$30~MeV. 
The higher-energy region, however, observes a significant increase due to a large contribution from the BH formation. 
In addition, our model shows an enhancement at low energies below $\sim$10~MeV compared to previous models. 
This feature is considered to be the result of redshifted neutrinos from $z>1$, where the core-collapse rate is higher because of a larger contribution from the early-type galaxies\footnote{A similar enhancement at low energies is pointed out by nonstandard physics, such as neutrino nonradiative decay, in \citet{2023PhRvD.107b3017I}.}. 
Figure~\ref{fig:fluxintegcomp} shows the integrated DSNB $\bar\nu_e$ fluxes with different combinations of SFR, EOS, and neutrino mass hierarchy for the reference and alternative models. 
Here the flux integrations are performed for two energy regions, $13.3\leq E_{\nu}<17.3$~MeV (low) and $17.3\leq E_{\nu}<31.3$~MeV (high), to visualize different features between the reference model and the others. 
These energy regions are also used in the later sensitivity analysis, in Section~\ref{sec:sens}. 
We find that BHs primarily contribute to the flux at high energies, while the flat IMF of the early-type galaxies enhances the flux broadly.

\section{DSNB Detection} \label{sec:exp}

We consider the DSNB detection at water-based Cherenkov detectors, especially the cases at SK and its upgrades, SK-Gd~\citep[][]{2022NIMPA102766248A} and HK~\citep[][]{2018arXiv180504163H}. 
SK-Gd is a new phase of SK with gadolinium loaded, which started in 2020. 
HK is a new detector with an $\sim$8.4 times larger volume than SK that is planned for construction. 
At water Cherenkov detectors, the main detection channel of the DSNB is the inverse beta decay (IBD) of electron antineutrinos, $\bar\nu_e + p \to e^+ + n$, as its cross section is largest in the region of a few tens of megaelectronvolts where the experimental search is performed. 
Here the final-state neutron has to somehow be converted to electromagnetic particles to be observed, while the positron can be detected easily via Cherenkov photons. 
This is achieved through the capture of a neutron after thermalization in water; a 2.2~MeV $\gamma$-ray is emitted after a neutron is captured on hydrogen, or multiple $\gamma$-rays whose energies total $\sim$8~MeV are emitted for the case of neutron capture on gadolinium. 
Only neutron capture on hydrogen is possible at SK and HK, while those on both hydrogen and gadolinium can occur at SK-Gd. 
The assumed neutron tagging efficiency at SK-Gd is 70\% in the following part, which is corresponding to a 0.02\%--0.03\% doping rate. 
For HK, we assume the same tagging efficiency as SK.

\begin{figure}[htbp]
  \gridline{\fig{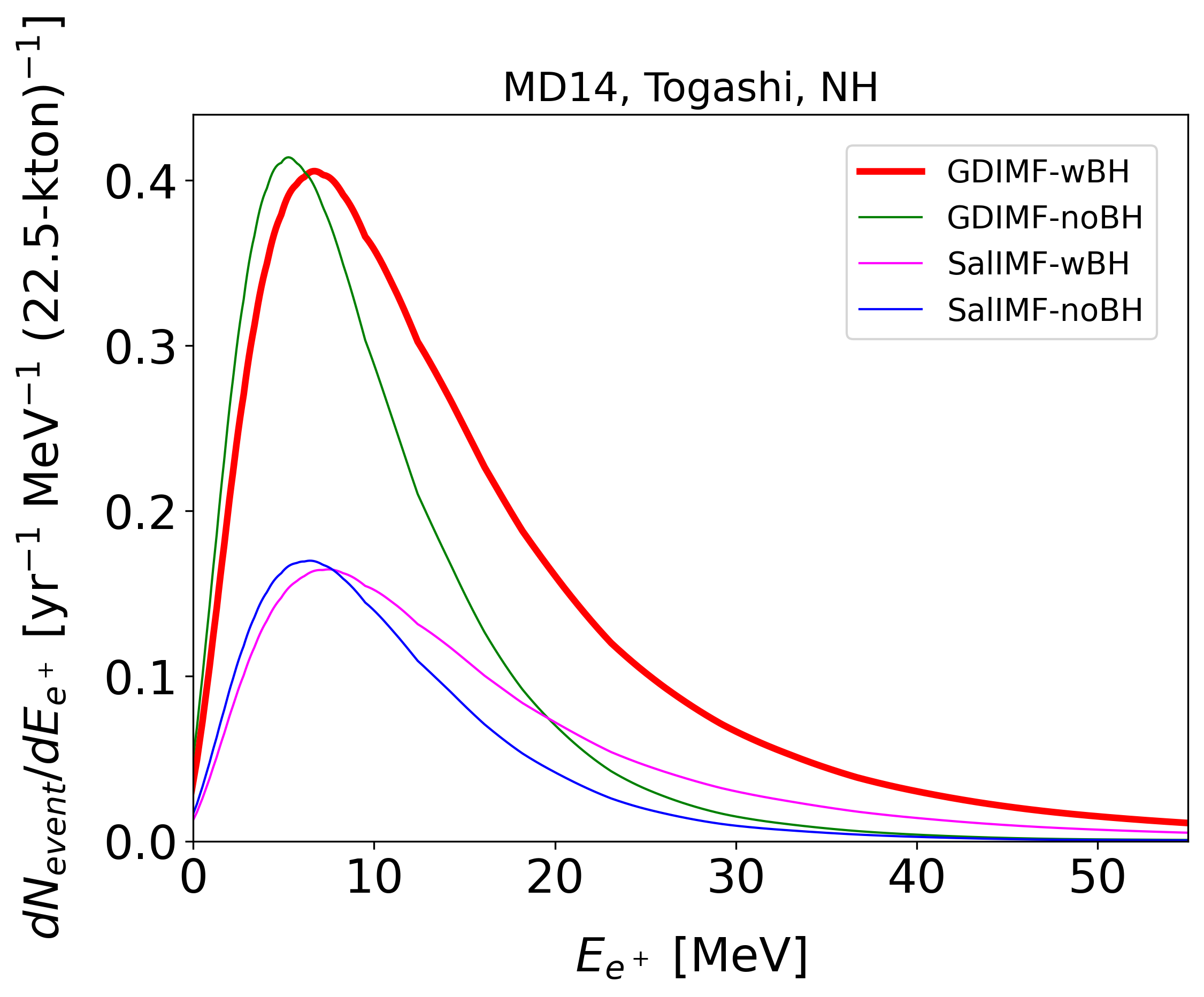}{0.45\textwidth}{}}
  \vspace{-22truept}
  \gridline{\fig{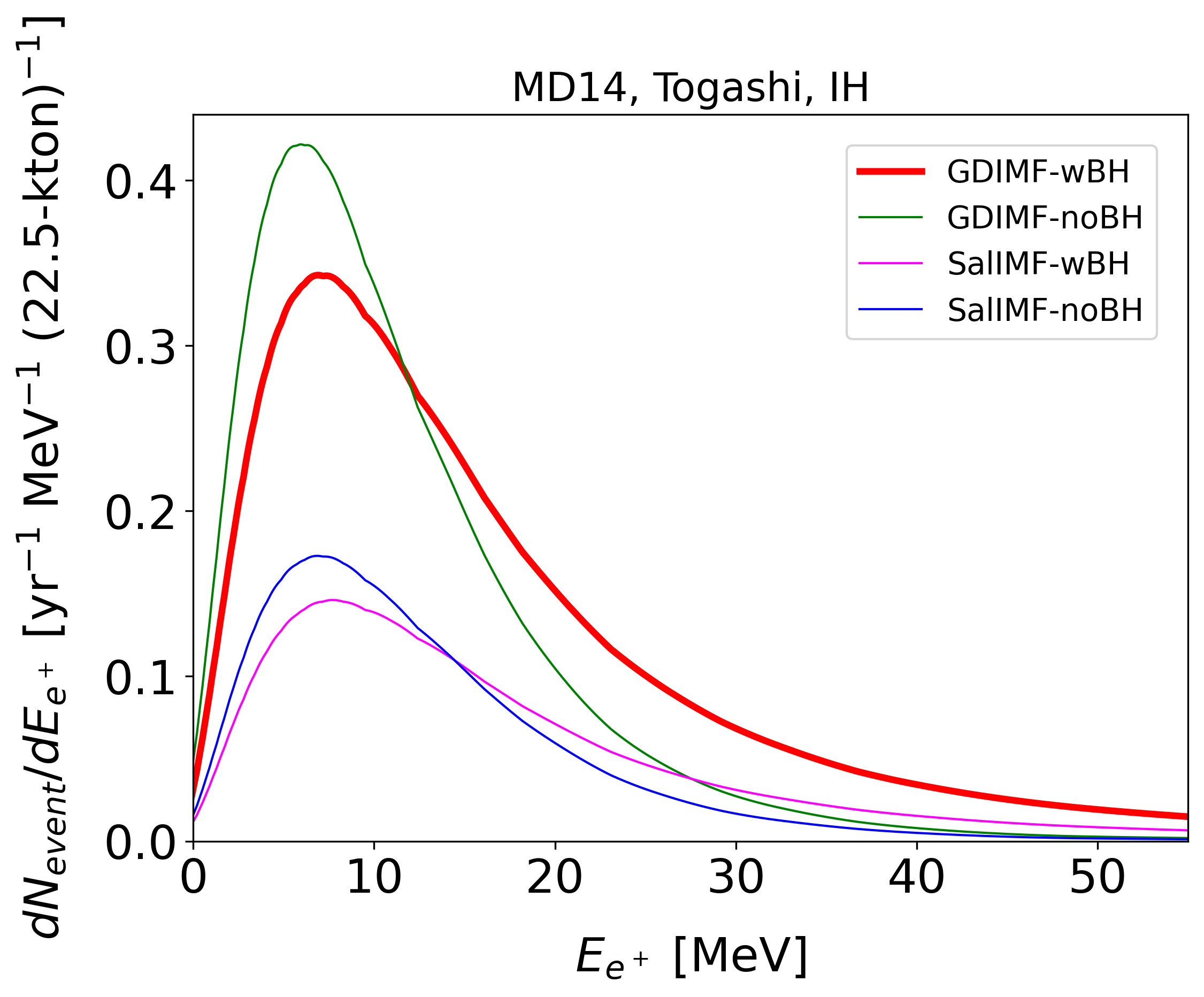}{0.45\textwidth}{}}
  \vspace{-12truept}
  \caption{Event rate spectra of the DSNB $\bar\nu_e$ at a water volume per 22.5~kton per yr with different choices of the IMF and BH treatment for NH (top) and IH (bottom). Here, MD14 and Togashi are assumed as the SFR and EOS, respectively. The thick red lines correspond to our reference model.}
  \label{fig:neventexample}
  \end{figure}

Figure~\ref{fig:neventexample} shows event rate spectra from our models calculated as the $\bar\nu_e$ flux times the IBD cross section in \citet{2003PhLB..564...42S}\footnote{The IBD cross section calculation has been updated recently in \citet{2022JHEP...08..212R}, but we perform the analysis with \citet{2003PhLB..564...42S}, as their difference is not sizable enough to change the present result.}. 
A variable IMF increases the flux in general, and the BH formation enhances the high-energy side. 
These features are common for the NH and IH cases. 
From our reference model (GDIMF-wBH) with the MD14 SFR and Togashi EOS, the expected numbers of IBD signal events at SK-Gd and HK over 10~yr are 15--18 and 50--60 (depending on the neutrino mass hierarchy), respectively.

\section{Experimental Sensitivity} \label{sec:sens}

We estimate the experimental sensitivity to our new DSNB model. 
Here, we consider water-based Cherenkov detectors, SK-Gd and HK.

\subsection{Background Sources}

In the water Cherenkov detector, requiring a combination of the positron and the $\gamma$-ray(s) from thermal neutron capture enables searches for a subtle signal of the DSNB $\bar\nu_e$ among a huge amount of background events, but there are still multiple types of remaining background. 
Background sources in the range of 10--100~MeV include atmospheric neutrinos, cosmic-ray muon spallation, and solar neutrinos.
In \citet{2021PhRvD.104l2002A}, the atmospheric neutrino background is categorized into two: the neutral-current quasielastic (NCQE) interaction, and the others (non-NCQE), including the charged-current interaction and the NC pion production, etc. 
NCQE can leave the $\gamma$-ray(s) and neutron(s) in their final states, thus mimicking the IBD signal~\citep[][]{2019PhRvD..99c2005W,2019PhRvD.100k2009A}. 
Charged-current and pion production interactions can have an invisible muon with a low momentum, which decays into a visible positron or electron and neutron(s), which then become background for the $\bar\nu_e$ search. 
Some of the energetic muons break up oxygen nuclei inside the detector and produce secondary hadronic particles, which react with other nuclei to further produce radioactive isotopes. 
These isotopes decay with a $\beta$ emission and some of them accompany a neutron. 
Among such $\beta$+$n$ decay isotopes, ${\rm ^{9}Li}$ is the most problematic in the DSNB search because of its high production rate, $\beta$ energy, and long lifetime~\citep[][]{2016PhRvD..93a2004Z,2021PhRvD.104l2002A,2021arXiv211200092S}. 
A positron-like signal from many other spallation isotopes or electrons by solar neutrino interactions may be detected accidentally at the same timing as a $\gamma$-like signal from low-energy events (e.g., photomultiplier noise). 
This accidental coincident background is sizable and then impairs the signal efficiency at SK and HK, while it can be rejected effectively at SK-Gd, due to the higher energies of the $\gamma$-ray signal. 
We follow \citet{2022ApJ...937...30A} on background properties and detector performance. 
Since both SK-Gd and HK have a lot in common with SK, the background compositions and their systematic uncertainties are extrapolated from the latest SK analysis~\citep[][]{2021PhRvD.104l2002A,2022ApJ...937...30A}.

\subsection{Analysis Procedure}

\begin{figure*}[htbp]
  \gridline{\fig{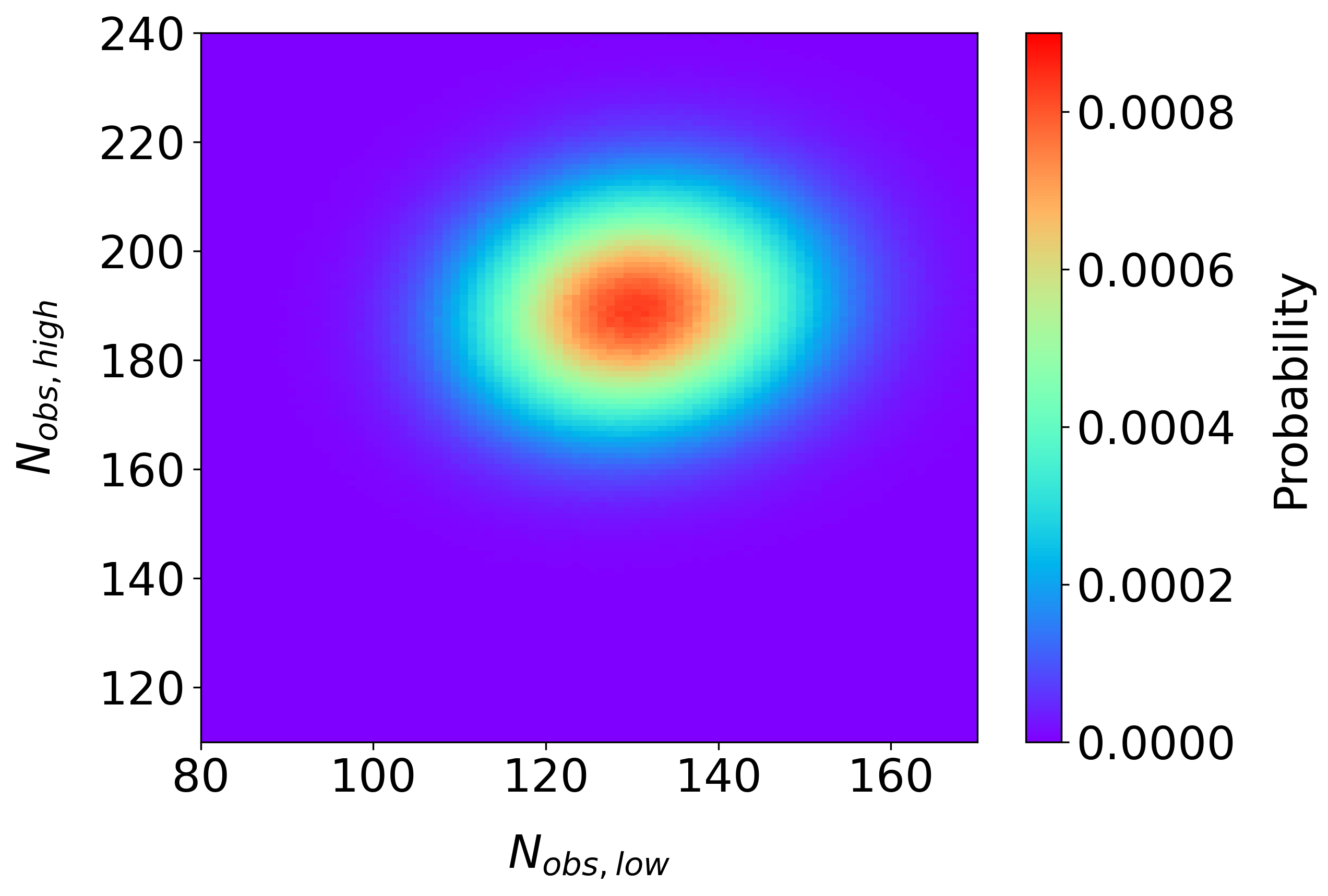}{0.46\textwidth}{(1) GDIMF-wBH}\hspace{-15truept}
            \fig{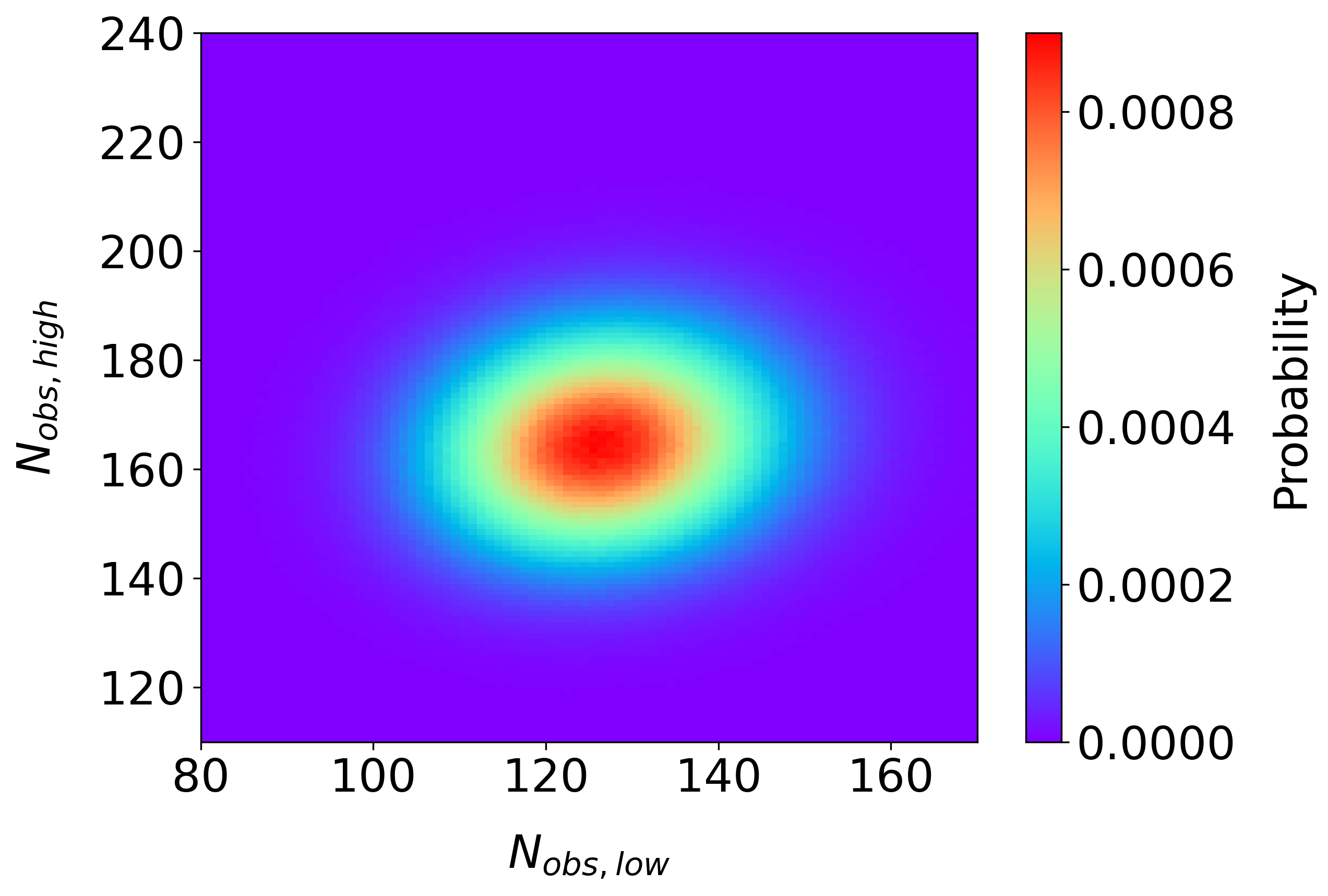}{0.46\textwidth}{(2) GDIMF-noBH}}
  \gridline{\fig{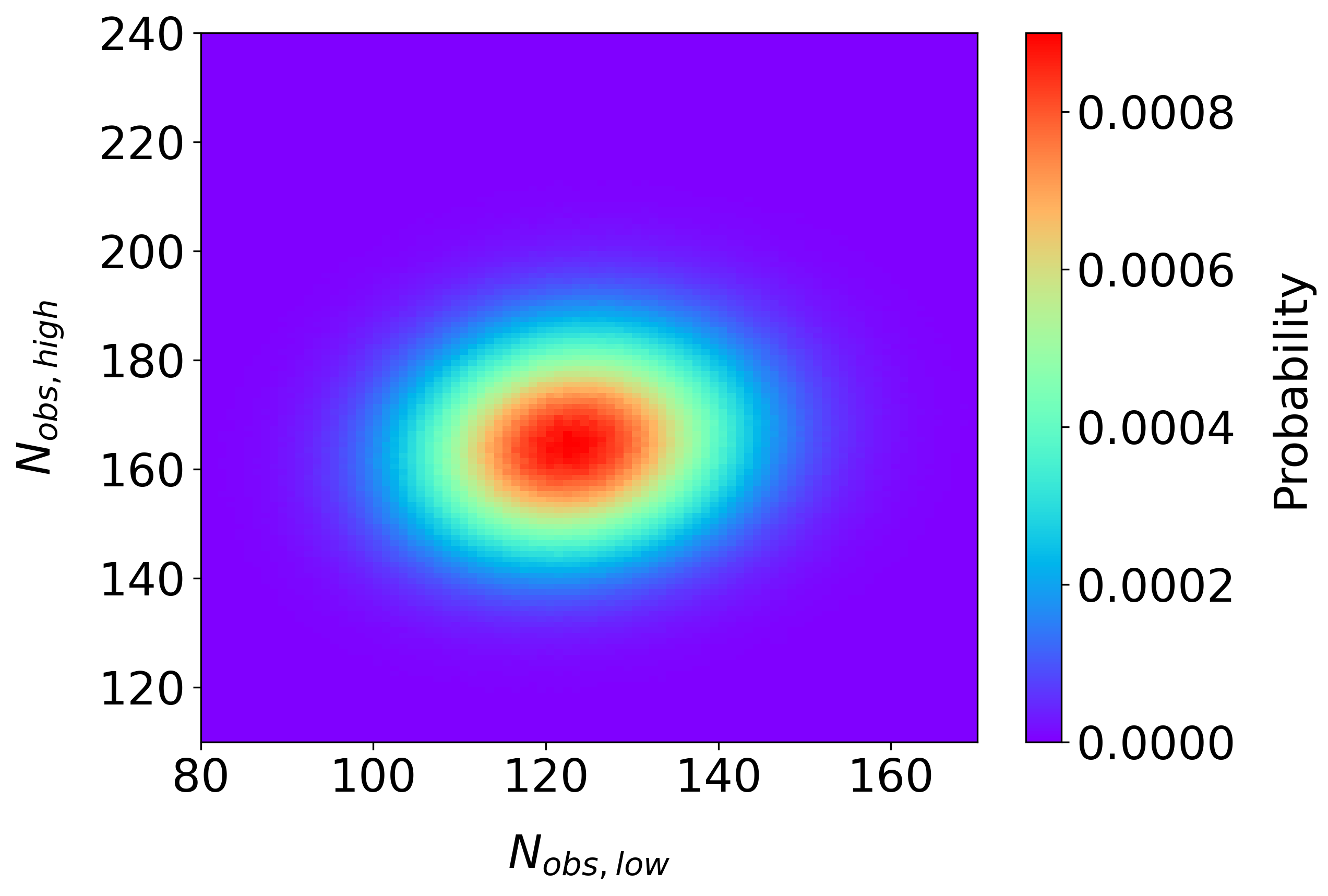}{0.46\textwidth}{(3) SalIMF-wBH}\hspace{-15truept}
            \fig{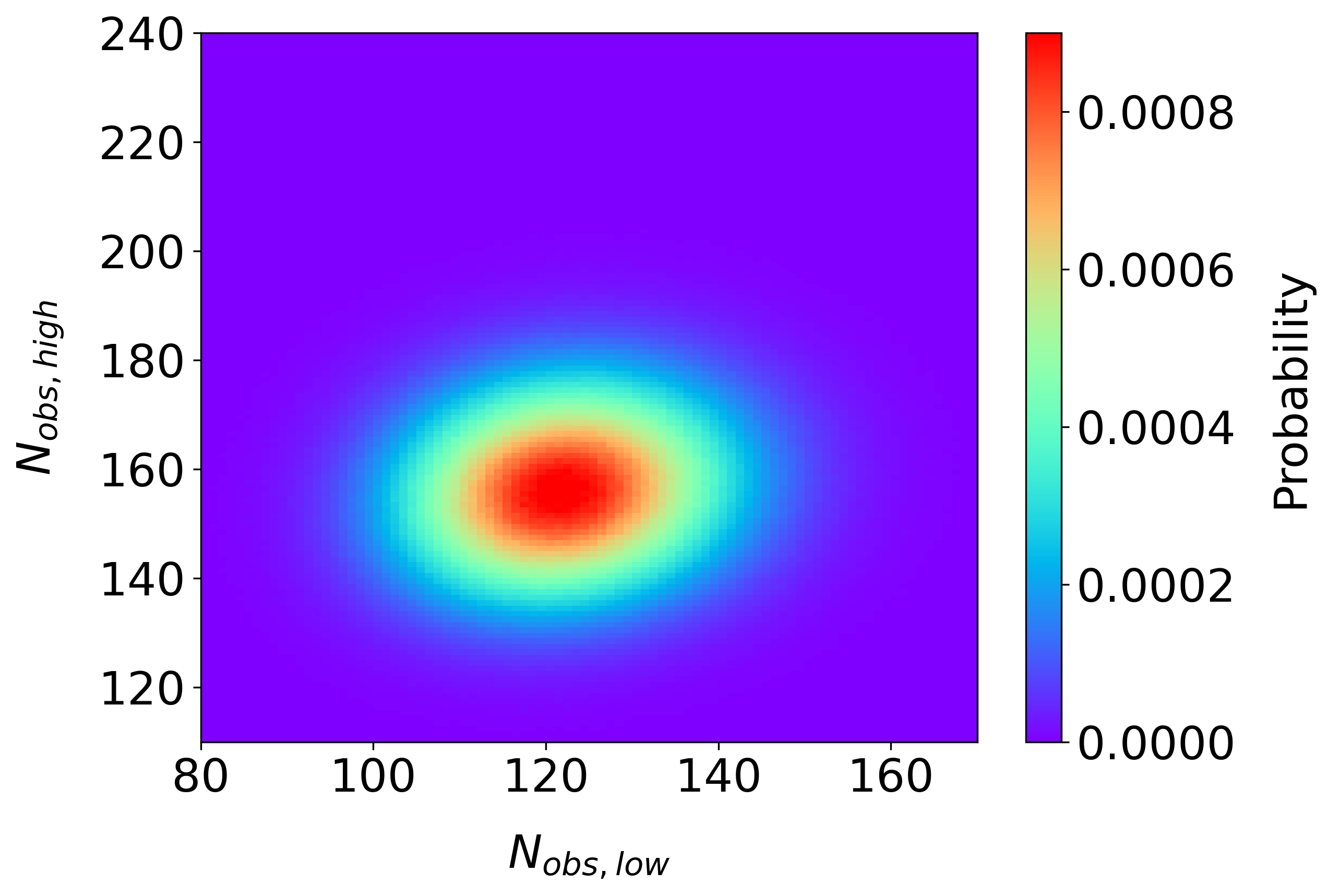}{0.46\textwidth}{(4) SalIMF-noBH}}
  \vspace{+5truept}
  \caption{2D normalized histograms (likelihood) on $N_{\rm obs,low}$ and $N_{\rm obs,high}$ for the case of the MD14 SFR, Togashi EOS, and NH at HK over 10 yr. Each panel corresponds to the models in Table~\ref{tab:modelnomenclature}.}
  \label{fig:likelihood2dexample}
  \end{figure*}

We try to extract constraints on DSNB models based on observation. 
This is achieved by calculating the posterior probability of a certain model with input data using Bayes' theorem, as
  \begin{eqnarray}
    P({\rm model}|{\rm data}) 
    = \frac{P({\rm data}|{\rm model}) \times P({\rm model})}{P({\rm data})} \nonumber \\ 
    = \frac{P({\rm data}|{\rm model}) \times P({\rm model})}{\displaystyle \sum_{\rm model} P({\rm data}|{\rm model}) \times P({\rm model})},
  \label{eq:bayes}
  \end{eqnarray}
where $P({\rm model})$ and $P({\rm data}|{\rm model})$ represent the prior probability and likelihood based on a model, respectively. 
The sum in the denominator is performed over all of the considered models. 
We use a uniform prior, i.e., $1/N$ as a prior, when testing $N$ models. 
In this study, we make use of the numbers of events in two energy ranges, $N_{\rm obs,low}$ for $13.3\leq E_{\nu}<17.3$~MeV and $N_{\rm obs,high}$ for $17.3\leq E_{\nu}<31.3$~MeV, as observables from data. 
This helps to discriminate models better than when using only one observable, as is indicated in Figure~\ref{fig:fluxintegcomp}. 
The number of observed events is the sum of the signal and backgrounds ($N_{\rm obs}=N_{\rm sig}+N_{\rm bkg}$) in each energy range. 
Example 2D likelihood distributions at HK over 10 yr are shown in Figure~\ref{fig:likelihood2dexample}. 
So as to evaluate the posterior probability in Equation~\ref{eq:bayes}, as input data, we use the real data from the fourth phase of SK (SK-IV), $\{N_{\rm obs,low},N_{\rm obs,high}\}=\{20,9\}$~\citep[][]{2021PhRvD.104l2002A}, for the result of SK, and the nominal expectations from a certain model for the sensitivities of SK-Gd and HK. 

\subsection{Signal Significance}

\begin{figure*}[htbp]
  \begin{center}
    \includegraphics[clip,width=11.5cm]{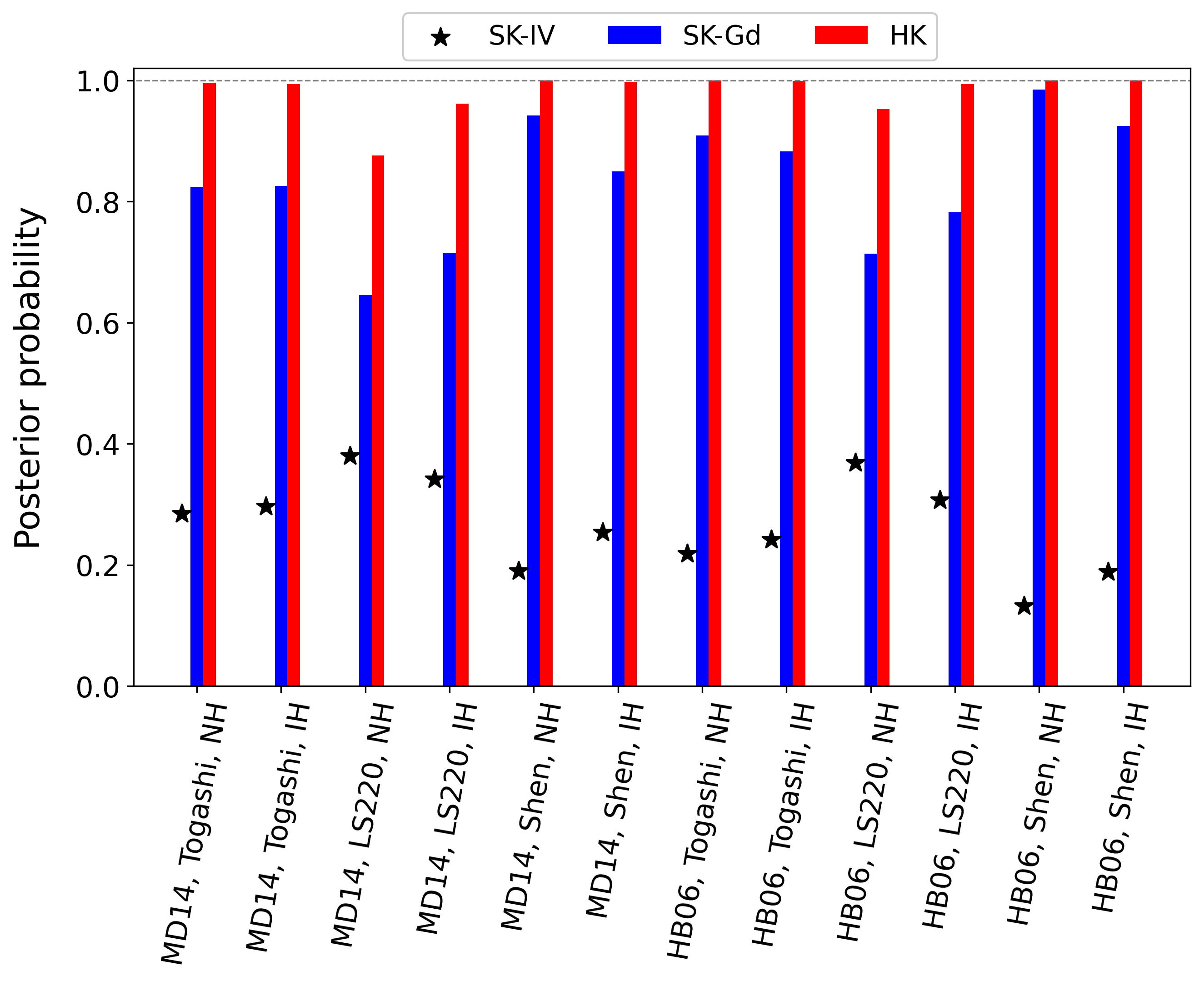}
  \end{center}
  \vspace{-10truept}
  \caption{Posterior probabilities of discriminating the reference DSNB model (GDIMF-wBH) against the background-only case at SK-IV, SK-Gd, and HK. As denoted by the labels, different SFR, EOS, and neutrino mass hierarchy cases are tested. Real data are used for SK-IV, while the nominal expectations from each model plus background are used for SK-Gd and HK as inputs.}
  \label{fig:discprobbkgonly}
  \end{figure*}

First, we test our DSNB model against the case of the background only (no DSNB signal) to investigate the significance of the model. 
Here we compare the reference model (GDIMF-wBH) and the background-only case by taking 1/2 as a prior. 
This is repeated for different choices of SFR, EOS, and neutrino mass hierarchy. 
The resulting posteriors are shown in Figure~\ref{fig:discprobbkgonly}.
The results for SK-IV are based on the real data from \citet{2021PhRvD.104l2002A} and imply more likeliness of the background only, although it is not significant yet. 
The results for SK-Gd and HK are obtained by injecting the nominal expectations from each DSNB model plus background as input data. 
Some models can be well discriminated from the background-only case at SK-Gd and even most models can be at HK. 
The least significant model is based on the MD14 SFR, LS220 EOS, and NH, which can be understood by the smallest $N_{\rm obs,low}$ and $N_{\rm obs,high}$ as seen in Figure~\ref{fig:fluxintegcomp}.
The posterior for this model is the largest from the SK-IV observation, since it is most similar to the background only. 

\subsection{Model Discrimination}

\begin{figure}[htbp]
  \begin{center}
    \includegraphics[clip,width=8.6cm]{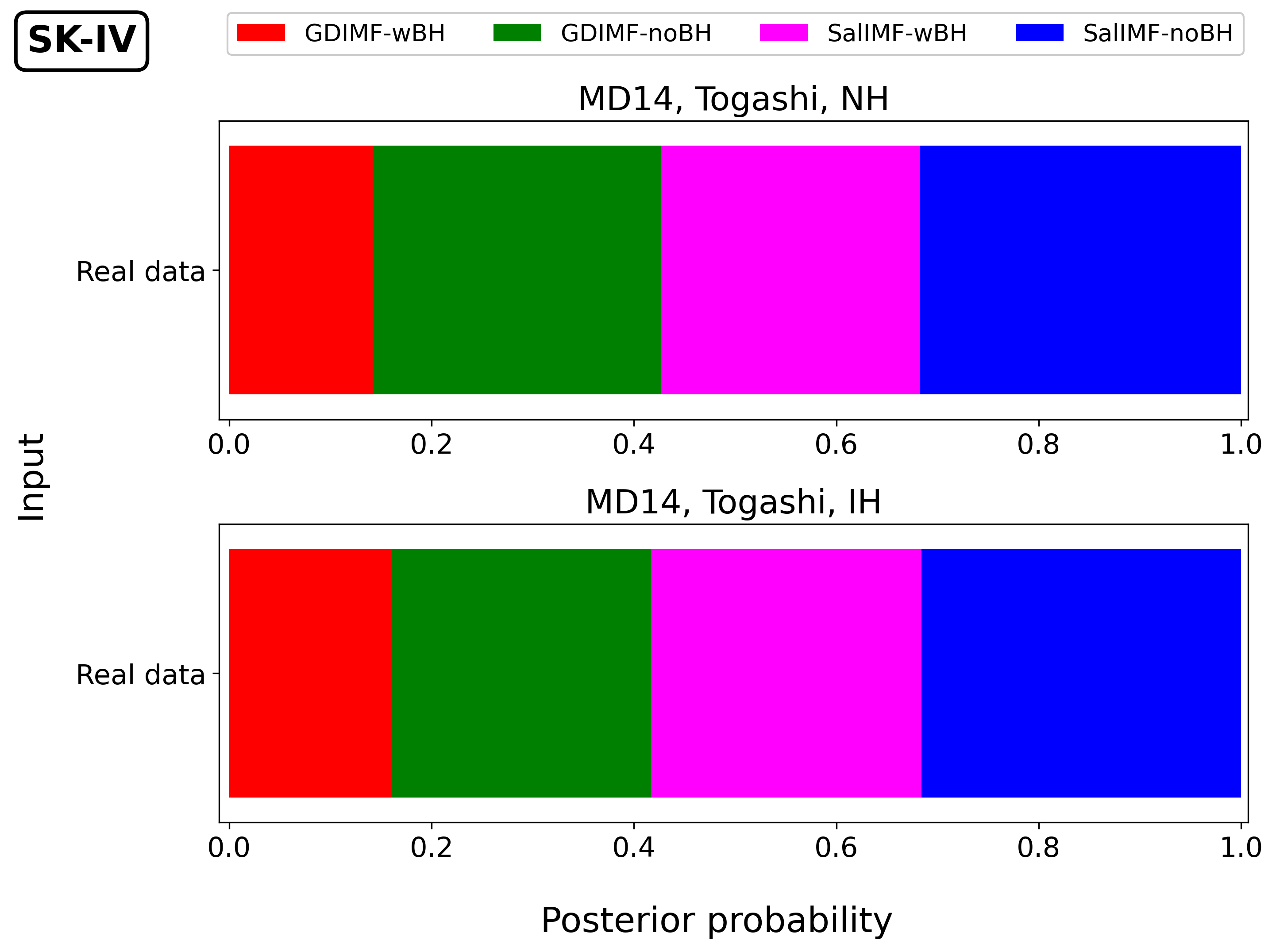}
  \end{center}
  \vspace{-5truept}
  \caption{Probabilities of discriminating the reference and alternative DSNB models with the MD14 SFR and Togashi EOS based on the SK-IV observation result. The top and bottom correspond to the NH and IH cases, respectively.}
  \label{fig:posteriorhb06togashisk4}
  \end{figure}

\begin{figure*}[htbp]
  \gridline{\fig{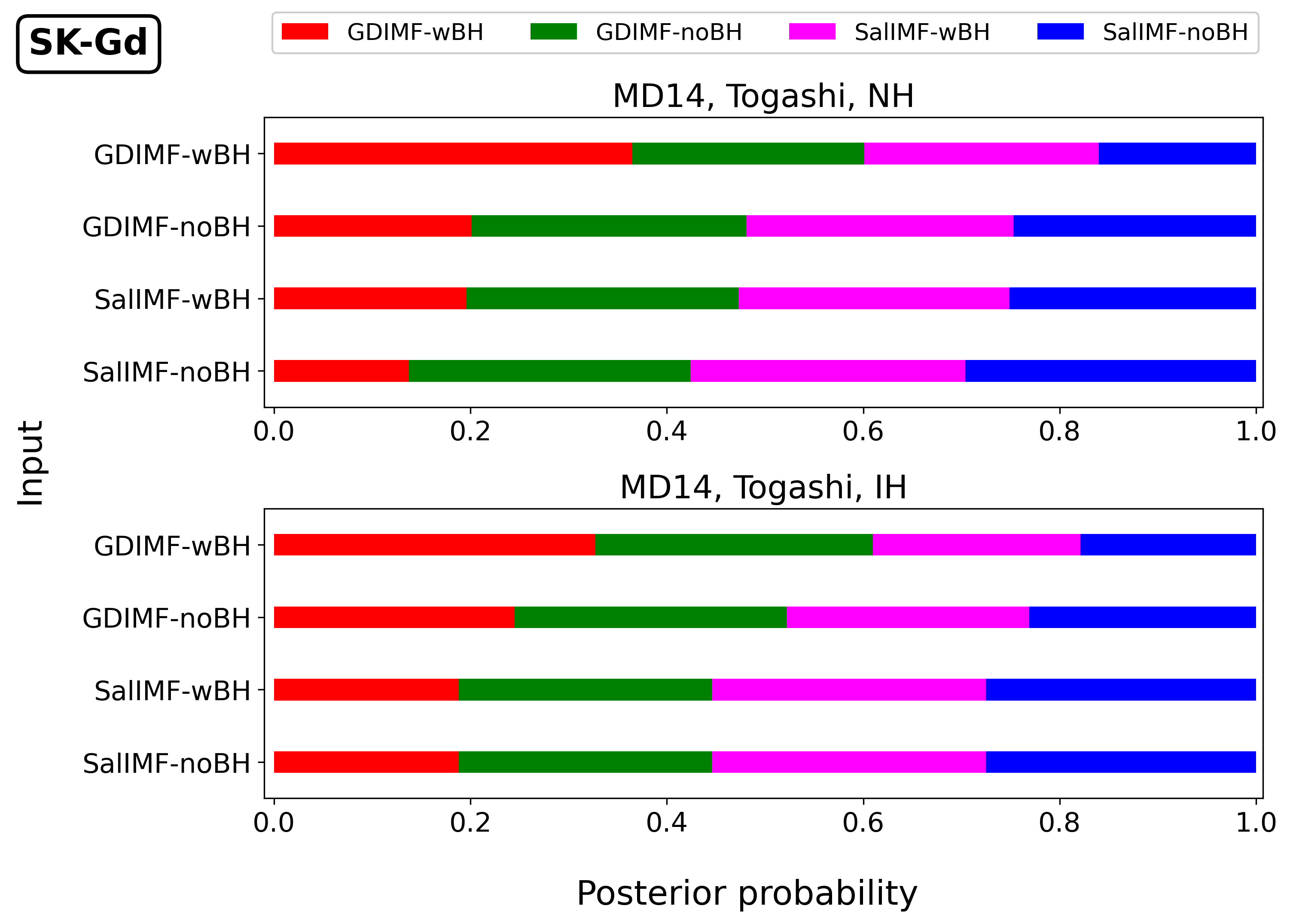}{0.48\textwidth}{}\hspace{0truept}
            \fig{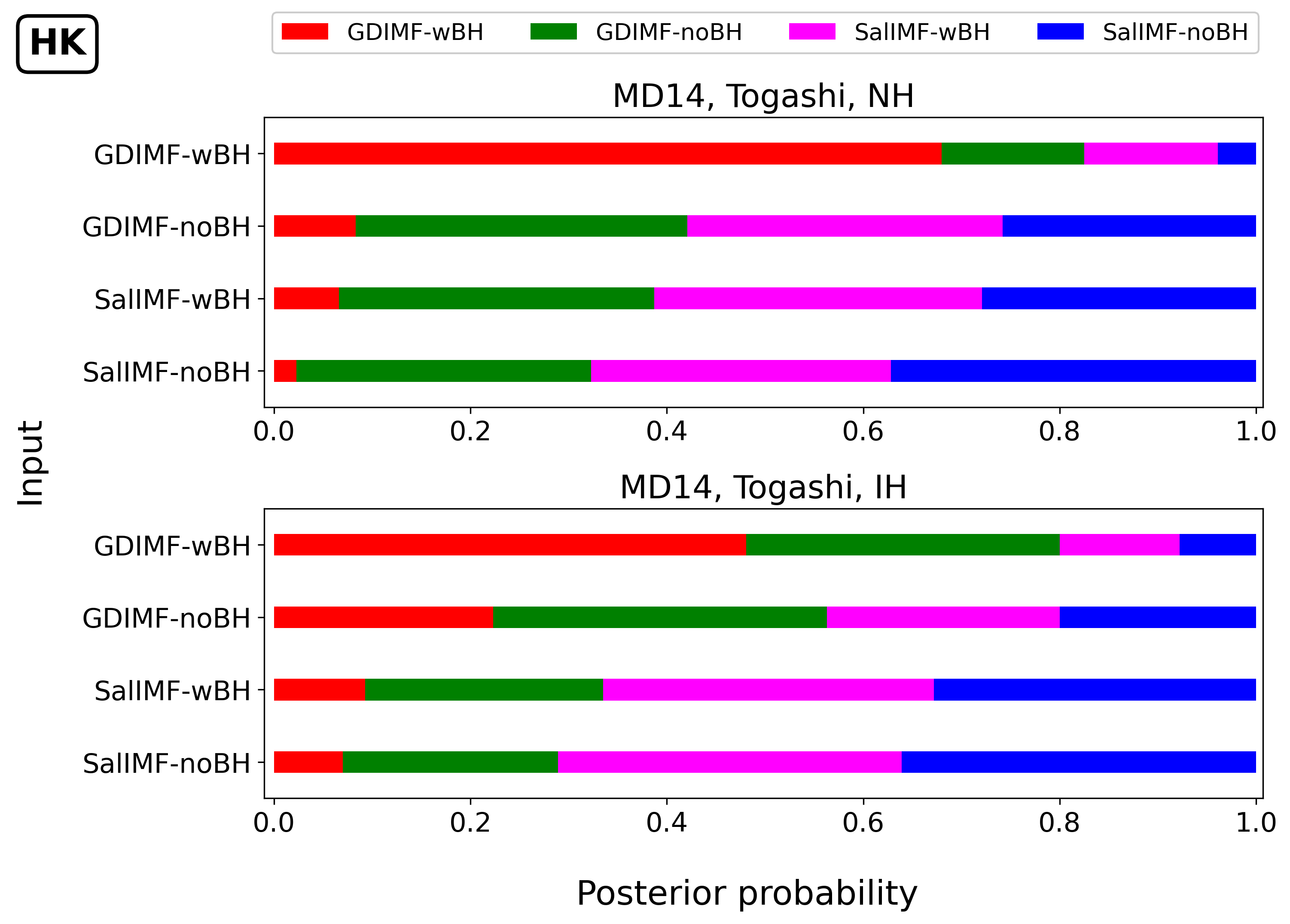}{0.48\textwidth}{}}
  \vspace{-12truept}
  \caption{The same graphs as Figure~\ref{fig:posteriorhb06togashisk4} but for SK-Gd (left) and HK (right). As explained in the main text, these are based on injections of the nominal expectations from each model plus the background as input data. Each row corresponds to the result with an input from each model in Table~\ref{tab:modelnomenclature}.}
  \label{fig:posteriormd14togashiskgdandhk}
  \end{figure*}

We then test our reference model (GDIMF-wBH) against the alternative models in Table~\ref{tab:modelnomenclature} for the purpose of investigating the impacts of the assumptions in the new model.
The posterior probabilities calculated based on the SK-IV data with the MD14 SFR and Togashi EOS are shown in Figure~\ref{fig:posteriorhb06togashisk4}. 
From the SK-IV data, the most (least) favored model among the four is SalIMF-noBH (GDIMF-wBH), especially because of a smaller $N_{\rm obs,high}$ in observation than expectation, while it is not significant enough. 
The sensitivity results for SK-Gd and HK with the MD14 SFR and Togashi EOS are shown in Figure~\ref{fig:posteriormd14togashiskgdandhk}. 
It is found that HK has a much stronger discrimination capability than SK-Gd. 
Therefore, we focus on the HK results in the following part.
In the NH case, the reference model can be discriminated well from the alternative models because of a significantly larger flux in both energy regions (see the top of Figure~\ref{fig:neventexample}). 
On the other hand, in the IH case, since the relative intensity of the DSNB flux from GDIMF-wBH to that from GDIMF-noBH is lower than the NH case (Figure~\ref{fig:neventexample}), the separation of these two models is more difficult. 
Nevertheless, sensitivity to the IMF assumption (variable or Salpeter) still looks good at HK (compare GDIMF-wBH$+$GDIMF-noBH vs. SalIMF-wBH$+$SalIMF-noBH in the bottom of the right panel in Figure~\ref{fig:posteriormd14togashiskgdandhk}). 

\begin{figure*}[htbp]
  \gridline{\fig{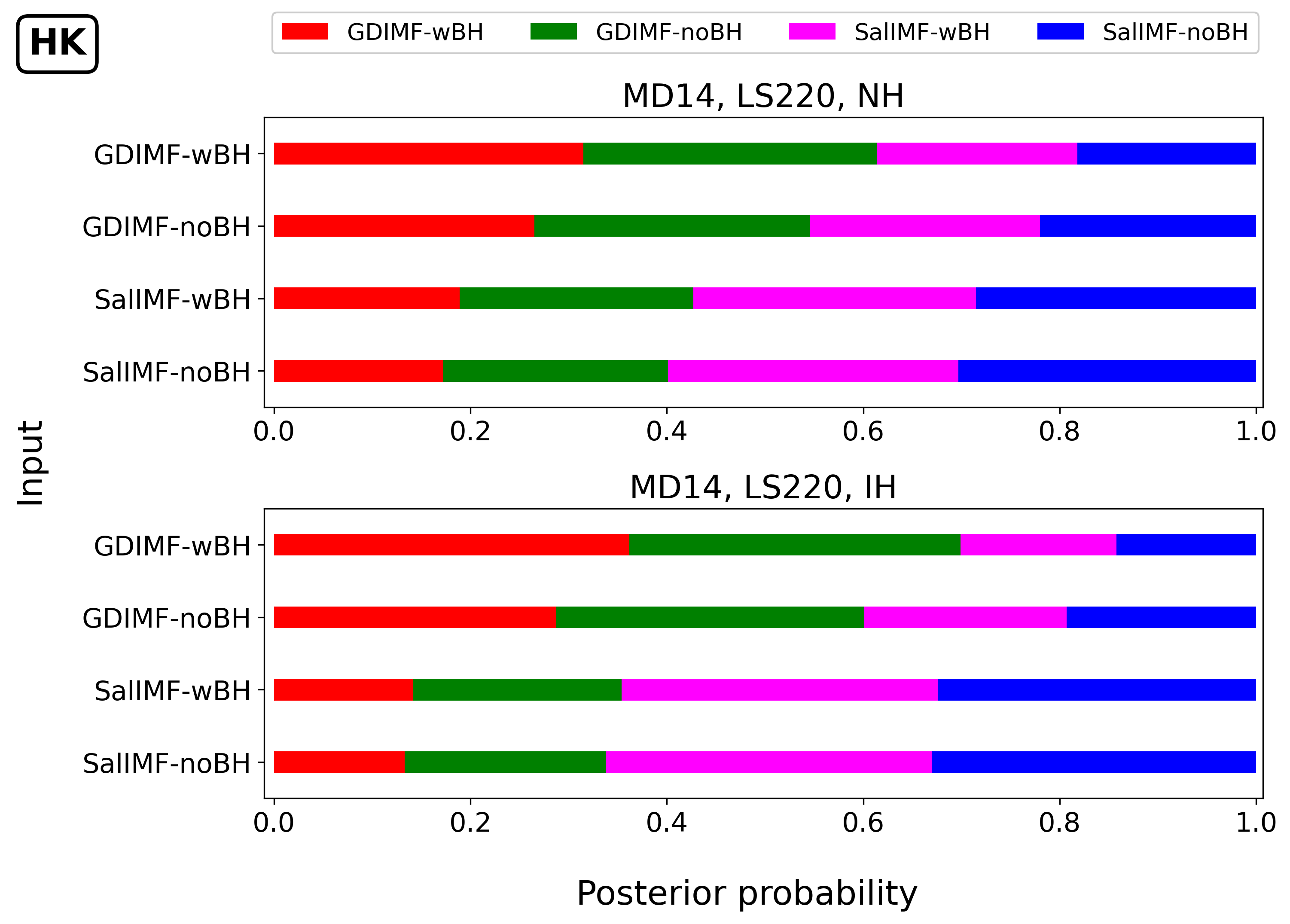}{0.48\textwidth}{}\hspace{0truept}
            \fig{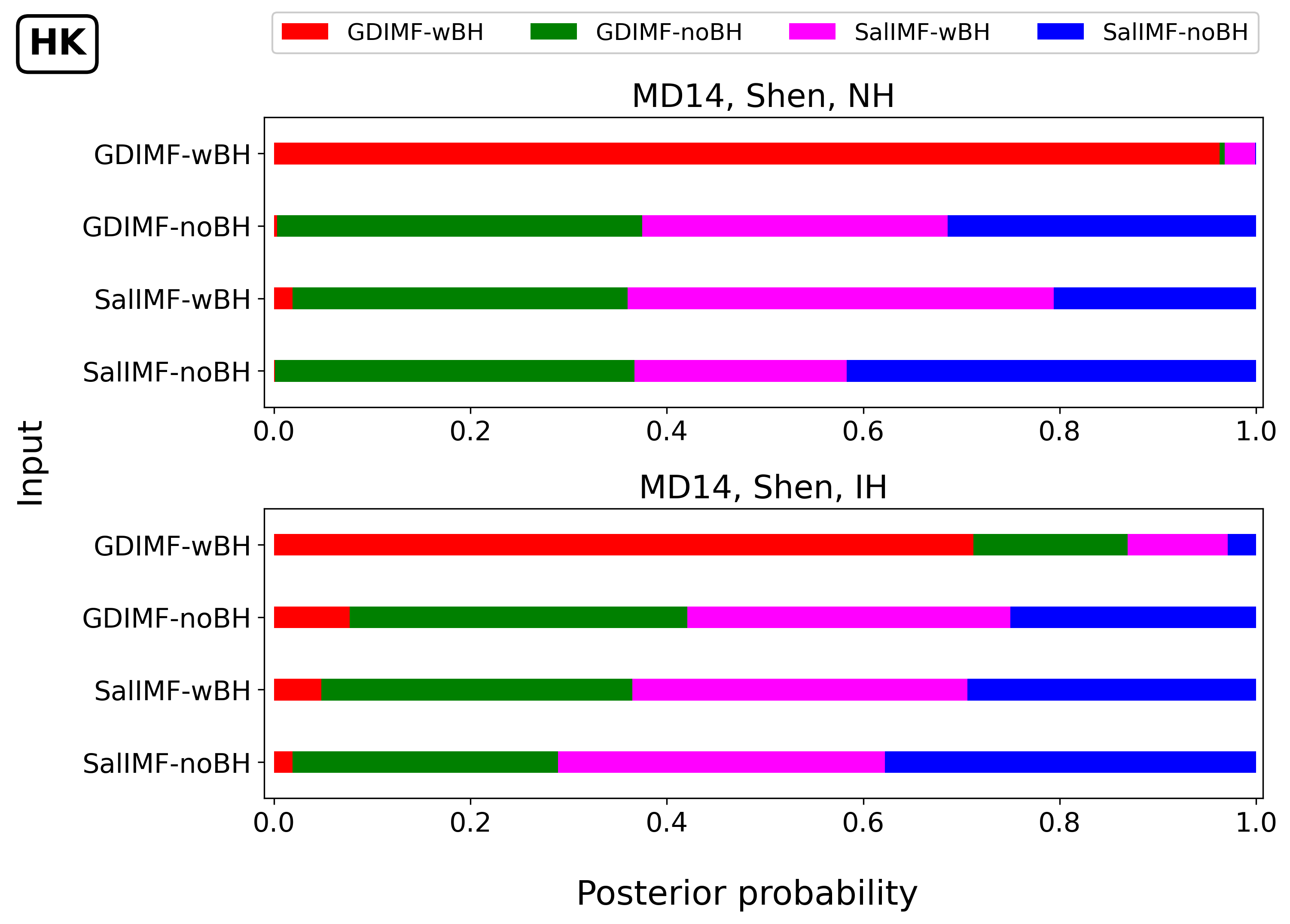}{0.48\textwidth}{}}
  \vspace{-12truept}
  \caption{The same graphs as Figure~\ref{fig:posteriormd14togashiskgdandhk} but for the LS220 (left) and Shen EOS (right) at HK.}
  \label{fig:posterioreosdep}
  \end{figure*}

\begin{figure}[htbp]
  \begin{center}
    \includegraphics[clip,width=8.6cm]{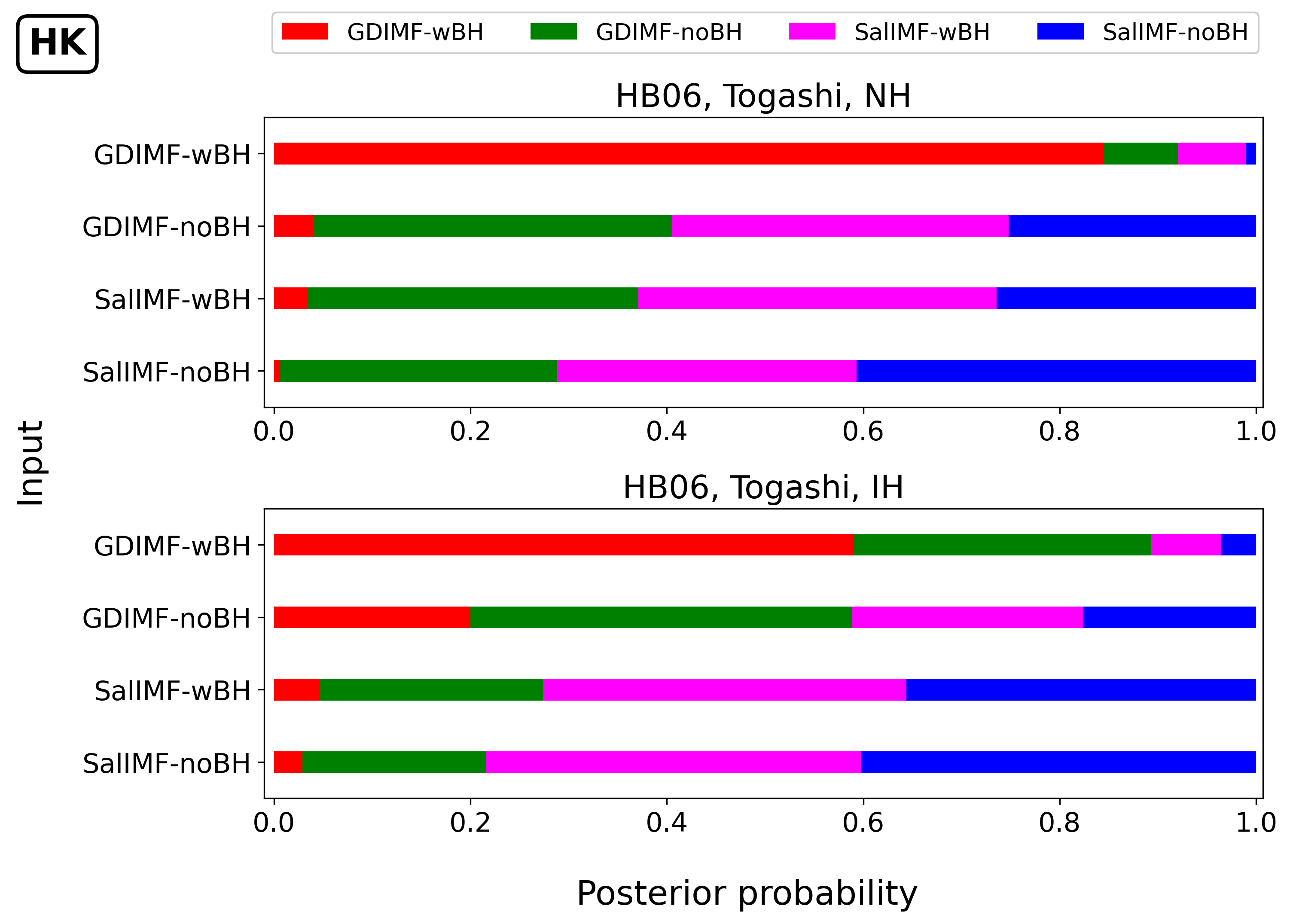}
  \end{center}
  \vspace{-5truept}
  \caption{The same graphs as Figure~\ref{fig:posteriormd14togashiskgdandhk} but for the HB06 SFR at HK.}
  \label{fig:posteriorsfrdep}
  \end{figure}

The sensitivities at HK with different choices of the nuclear EOS, LS220 or Shen, are shown in Figure~\ref{fig:posterioreosdep}. 
The contribution from the BH formation is smallest with LS220, while it is largest with Shen, as explained by their maximum mass of NSs (see Section~\ref{sec:model}). 
This is strongly related to how well the reference model can be discriminated from the alternative ones. 
The results in Figure~\ref{fig:posterioreosdep} clearly show that the separation of GDIMF-wBH from the others is hard with LS220, while almost perfectly possible with Shen. 
However, again, separation of the IMF type looks possible even in the LS220 case. 
The results with the HB06 SFR and Togashi EOS are also shown in Figure~\ref{fig:posteriorsfrdep}. 
Comparing this and the right panel of Figure~\ref{fig:posteriormd14togashiskgdandhk}, model discrimination is easier with HB06, because the overall flux is larger than the case of MD14, due to the higher SFR. 

\section{Discussion} \label{sec:discuss}

In this study, we assume a variable IMF, which generates more numerous massive stars, with a negligible change in the number of intermediate-mass stars such as $\sim$3$M_\odot$ in a higher-SFR environment. 
Such SFR-dependent IMFs can reproduce the observed break in the proportionality relating cosmic star formation to the CCSN rate for $0<z<1$ \citep{2023MNRAS.518.3475T}. 
In contrast, it is shown that an IMF whose slope varies, including intermediate-mass stars, is not in good agreement with the cosmic CCSN rate, owing to a resultant low cosmic star formation \citep{2022MNRAS.517.2471Z}. 
We stress that our adopted IMF is also consistent with observational evidence for the variable flux ratio of H$\alpha$ to UV emission, which trace the stars with $M$\gtsim 20$M_\odot$ and \gtsim 3$M_\odot$, respectively, among galaxies whose SFRs are different \citep{2009ApJ...695..765M, 2009ApJ...706..599L}. 
However, we still need more efforts in the future to connect cosmic star formation to the CCSN rate in terms of nonuniversal IMFs to raise the accuracy of the DSNB flux prediction. 
In addition, we must keep in mind that there could be an uncertainty in the predicted frequency of BH formation against redshift; there are implications that the high-mass end of the IMF might correlate with the magnitude of the SFR \citep[e.g.,][]{2009ApJ...695..765M,2009ApJ...706..599L,2009MNRAS.395..394P}.

As is mentioned in Section~\ref{sec:model}, CCSNe with a high-mass NS are expected to show different neutrino spectra than those with a canonical-mass NS. 
In \citet{2022ApJ...937...30A}, neutrino spectra from these two models are mixed with a fraction of CCSNe with a high-mass NS to those with an NS. 
Also, the released neutrino energy from the failed SN with a BH formation differs for various factors. 
For example, \citet{2021ApJ...909..169K} discuss the case that the BH formation with a delay of $\gg$1~second releases up to about twice larger energy of neutrinos than the fast case. 
We leave the detailed investigation of these systematic factors for the future work. 

While the current analysis focuses on the energy range between 13.3 and 31.3~MeV, there is a potential use of the outside region at future detectors. 
For example, HK is able to utilize the higher-energy region effectively because of its larger dataset. 
Also, liquid scintillators such as JUNO can lower the analysis threshold below the water Cherenkov detector and then potentially detect the DSNB at low energies below 13.3~MeV~\citep[][]{2022JCAP...10..033A,2022Univ....8..181L}. 
It should be noted that the usage of these high- and low-energy regions may be beneficial to identifying the DSNB signal from the present model, because significant enhancements in the flux are expected, as seen in Figure~\ref{fig:fluxmodelcomp}.
Further, better sensitivity can be expected with the potential doping of gadolinium to HK in the future.

\section{Conclusion} \label{sec:concl}

In this paper, we introduce a variable IMF, depending on the galaxy type, and consider the BH formation for progenitors with a mass larger than 18$M_{\odot}$ to calculate the DSNB flux. 
The calculations are performed for different combinations of the SFR (MD14 or HB06), EOS (Togashi, LS220, or Shen), and neutrino mass hierarchy (NH or IH). 
Our new model is characterized by the fact that the resultant $\bar\nu_e$ flux is enhanced at high and low neutrino energies compared to previous theoretical works. 
The enhancement at energies above $\sim$30~MeV is explained by a larger frequency of BH formation after stellar core collapse in this model. 
The flux increase at the lower-energy side ($\ltsim$10~MeV) is because of the feature that more core collapses are expected, due to a larger contribution from early-type galaxies, which leads to more neutrinos at the far Universe being redshifted. 
We show the event spectrum from our new model at the water Cherenkov detector and estimate the experimental sensitivities at SK-Gd and HK, in particular. 
It is found that the DSNB neutrinos from our model can be detected well over background events at these detectors.
We also prepare alternative models, based on the different IMF type and BH treatment, and show that the model discrimination between our model and these alternatives can be possible at HK.

\begin{acknowledgments}

This work is supported by Grants-in-Aid for Scientific Research (JP18H01258, JP20K03973, and JP23H00132) and a Grant-in-Aid for Scientific Research on Innovative Areas (JP19H05811) from the Ministry of Education, Culture, Sports, Science and Technology (MEXT), Japan. 

\end{acknowledgments}


\bibliography{sample631}{}
\bibliographystyle{aasjournal}



\end{document}